
\def\v{\varphi}

\def\t{\textstyle}


\vglue 2cm
\baselineskip=20pt
\centerline{\bf ON THE CLASSIFICATION OF SCALAR NON-POLINOMIAL}
\centerline{\bf EVOLUTION  EQUATIONS : QUASILINEARITY  }

\vskip 1cm
\baselineskip=14pt
\centerline {Ay\c se H\"umeyra Bilge}
\vskip .3cm

\leftline{ \it Department of
Mathematics, Istanbul Technical University, Maslak, Istanbul,
 Turkey} 
\leftline{
\it TUBITAK, Feza G\" ursey Institute, \c Cengelkoy, Istanbul, Turkey}

\vskip 1cm
\baselineskip=16pt
\centerline{\bf Abstract}

We prove that, for $m\ge 7$,  scalar evolution equations of the form
$u_t=F(x,t,u,\dots,u_m)$
which admit a nontrivial conserved density of order $m+1$ are linear
in $u_m$. The existence of such conserved densities is a necesary
condition for integrability in the sense of admitting a formal
symmetry, hence integrable scalar evolution equations of order $m\ge
7$ are quasilinear.

\vskip 1cm
\baselineskip=20pt
\noindent
{\bf 1. Introduction}

It has been shown that scale invariant  polynomial of evolution 
equations
of order greater than seven are symmetries of third and fifth order
equations [1] and similar results are  obtained in the
case where negative powers are involved [2].
The problem of classification of arbitrary evolution equations is thus
reduced to proving that such equations have desired polinomiality and
scaling properties. As a first step in this direction, 
we study scalar evolution equation of general type and 
show that integrable equations of the form
$$u_t=F(x,t,u,\dots,u_m)\eqno(1.1)$$
where $F$ has partial derivatives of all orders, is linear in 
$u_m$ for $m\ge 7$. 
The integrability criteria that we are using is the one given in
[3], namely the existence of a ``formal symmetry'' $R$, which is a
formal series in inverse povers of $D=d/dx$, satisfying the operator
equation 
$$R_t+[R,F_*]=0,\eqno(1.2)$$
where $F_*$ is the Frechet derivative of $F$.
The solvability of the
coefficients of $R$ in the class of local functions 
requires that certain quantities denoted as $\rho^{(i)}$ be conserved 
densities. 
These conserved density conditions give overdetermined systems of
partial differential equations for $F$ and lead to a classification.

For $m=3$, it is known that there are essentially nonlinear equations
that are candidates for being integrable [3] and their  study 
require special techniques as explored in [4].  Our proof
based on the existence of the  conserved density $\rho^{(1)}$ is not
applicable to third order equations, because for $m=3$, the expression
of  $\rho^{(1)}$ given by (2.11)  is no longer valid. 
For $m\ge 7$ the requirement that $\rho_1$ be a conserved density
leads to two  equations involving $\partial F/\partial u_m$ which  are 
are compatible only when $F$ is linear im $u_m$.  
For $m=5$ there are three 
such equations given  by (4.16a-c),  but these are compatible and we
cannot exclude the possibility of the existence of  essentially 
nonlinear equations at the fifth order.

In Section 2, we give  a concise overview  of the formal symmetry method and
present the expression of the conserved densities $\rho^{(i)}$,
$i=1,2,3$. In the proof of out quasilinearity result, we actually do
not need the explicit expression of these conserved densities, because
the only information we use is the existence of a conserved density of
order $m+1$ which is quadratic in $u_{m+1}$. 
The computation of the top order terms in derivatives are given in Section
3. Section 4 is devoted to general results and to the discussion of
third and fifth order equations. 
The main result is presented in Section 5.

\vskip 0.5cm
\noindent
{\bf 2. Notation and terninology, formal symmetries, conserved densities}

Let $u=u(x,t)$. A function $\v $ of  $x$, $t$, $u$ and the
derivatives of $u$ up to a fixed but finite order will be called a
``differential function'' [5] and  denoted by $\v[u]$. We
shall assume that $\v$ has partial derivatives of all orders. 
For notational convenience, we shall denote indices by subscripts  
or superscripts in
paranthesis such as in $\alpha_{(i)}$ or  $\rho^{(i)}$ and reserve 
subscripts without parantheses  
for partial derivatives, i.e.,
$$\eqalignno{
u=&u(x,t),\quad\quad 
    u_k={\partial^k u\over \partial x^k},
    \quad u_x={\partial u\over \partial x},\quad u_t={\partial u \over \partial t},\cr
\v=&\v(x,t,u,u_1,\dots,u_n),\quad\quad 
    \v_k={\partial \v\over \partial u_k},\quad 
    \v_x={\partial \v\over \partial x},\quad 
    \v_t={\partial \v \over \partial t}.
 &(2.1) \cr
}$$
In Section 5 only, we shall use the notation
    $\rho''=\rho_{m+1,m+1}$, 
in order to simplify the presentation. 
 
If $\v$ is a differential function, the total derivative with respect 
to $x$ is denoted by $D\v$ and it is
given by
$$D\v=\sum_{i=0}^n \v_i\ u_{i+1} \ + \ \v_x.\eqno(2.2)$$
Higher derivatives can be computed by appling the binomial formula as
$$ D^{k}\v= \sum_{i=0}^n\sum_{j=0}^{k-1}  \t{k-1 \choose j} D^j\v_i \ u_{i+k-j}
              \ +  \ D^{k-1}\v_x.
\eqno(2.3)$$
Note that even if $\v$ has an arbitrary functional form, $D^k\v$ is
polynomial in $u_{n+i}$ for $i\ge 1$. 
We shall denote generic functions that depend on at most $u_n$ by
$O(n)$. More precisely,
$$\varphi=O(u_n)\quad \quad 
{\rm if \ and \ only \ if }\quad  \quad 
\partial \v /\partial  u_{n+k}=0\quad {\rm for}\quad k\ge 1\eqno(2.4)$$
Clearly, if $\v=O(u_n)$, then $D\v$ is linear in $u_{n+1}$ and $D^k\v$
is polynomial in $u_{n+i}$ for $i\ge 1$.  
In certain places we shall need to know not only the order of a
function but whether it is a polynomial or not.  When this distinction is
important, 
 we shall use
the notation $\v=P(u_n)$, i.e., 
$$\varphi=P(u_n)\quad \quad 
{\rm if \ and \ only \ if }\quad  \quad 
\v=O(u_n)  \quad {\rm and }\quad 
\partial ^k \v /  \partial u_n^k=0
\quad {\rm for \ some }\ k.\eqno(2.5)$$
For $\v$ and $u$ as above, the total derivative with respect to $t$ is denoted
by $D_t\v$ and is given by
$$D_t\v=\sum_{i=0}^n\v_i D^iF\ + \ \v_t.\eqno(2.6)$$
Equalities up to total derivatives with respect to $x$  will be denoted by $\cong$, i.e.,
$$\varphi\cong \psi\quad\quad {\rm if \ and \ only \ if }\quad  \quad 
\varphi=\psi+D\eta,\eqno(2.7)$$
for some differential function $\eta$.

 A {\it symmetry}  $\sigma$ of $u_t=F[u]$  is a
differential function  which satisfies
the linearized equation $\sigma_t=F_*\sigma$ where $F_*$ is the Frechet
derivative of $F$ while a 
{\it conserved density} $\rho$  is a 
differential function  satisfying 
$\rho_t=D\eta$ for some  differential function $\eta$. The order of a
symmetry or of a conserved density is the order of the highest
derivative of $u$ in $\sigma $ or $\rho$.
The {\it recursion operator} $R$ is defined to be a linear operator that
sends symmetries to symmetries, i.e. $R\sigma$ is a symmetry whenever $\sigma
$ is a symmetry [5] and satisfies Eq.(1.2). 
A formal series which satisfies Eq.(1.2) up to a
finite order is called a {\it formal symmetry}.

In $1+1$ dimensions the integral terms in the  recursion operator can be
expanded in inverse powers of $D$ and one can look for a
formal series satisfying Eq.(1.2). 
 It is known that whenever $R$ is a solution, $R^{k/n}$ is also a 
solution of Eq.(1.2), hence  the  order of  $R$
 is irrelevant as long as we deal with recursion
operators expressed as a formal series in $D^{-1}$.

It is known that the quantities 
$$\rho^{(-1)}=F_m^{-1/m},\quad \quad\rho^{(0)}=F_{m-1}/F_m,\eqno(2.8)$$
where 
$$F_m={\partial F\over \partial u_m},\quad 
  F_{m-1}={\partial F\over u_{m-1}}\eqno(2.9)$$
are conserved densities for equations of any order [3].

For the computation of the conserved densities for arbitrary $m$, 
one can start with a
formal series  for $R$,  substitute in Eq.(1.2) and solve the operator
equation recursively, obtaining at each step linear first
order differential equations for the coefficients of $R$. The solvability of
these equations in the class of local functions gives the  conserved density
conditions, $\rho^{(i)}_t=D\eta^{(i)}$, including the ones above.

Alternatively, one can use the fact that the coefficient of $D^{-1}$ 
in the commutator of any two formal series is a total derivative [3].
If we start with a  formal series $R$  of order $1$ satisfying (1.2),
the coefficient of $D^{-1} $  will be the conserved
density $\rho^{(1)}$. By the remark above,  $R^k$ also satisfies
(1.2), hence the coefficient of $D^{-1}$ in $R^k$ will also be a
conserved density that we denote by $\rho^{(k)}$.  For $k=m$ the top
$m-2$ terms in $R^m$ should coincide with $F_*$, and the coefficients
of $R$ are expressed in terms of the partial derivatives 
of $F$ for each $m$.

For third order equations, the expression of $\rho^{(1)}$ involves
 $\eta^{(-1)}$ defined by
 $D_t\rho^{(-1)}=D\eta^{(-1)}$.  But for equations of  order $5$ or
 higher, $\rho^{(1)}$ and $\rho^{(2)}$ are independent of $\eta^{(-1)}$ and
 $\eta^{(0)}$.
Similarly for $5$th order equations, $\rho^{(3)}$ depends on $\eta^{(1)}$, but
 for $m\ge 7$, $\rho^{(3)}$ is independent of $\eta^{(i)}$ with $i\le 2$.  
We have computed  the  conserved densities $\rho^{(1)}$ and
$\rho^{(2)}$ for
$m\ge 5$ and $\rho^{(3)}$ for $m\ge 7$ with the method described above and
compared with the results of 
 direct integration of the first order equations obtained
recursively from Eq.(1.2). The formulas below were obtained
analytically for $m\ge 13$ and proved by induction. We have computed
the conserved densities at lower orders
directly using REDUCE and a symbolic integration package [6] and
checked that they obey the general formula. 
The results are given
in Eqs.(2.11-13) where we use the notation 
$$a=F_m^{1/m},\quad \alpha_{(i)}=F_{m-i}/F_m,\quad i=1,2,3,4.\eqno(2.10)$$
Then for $m\ge 5$ we have 
$$\eqalignno{
\rho^{(1)}=& a^{-1}(Da)^2
         -  {\t {12\over m(m+1)}} Da \ \alpha_{(1)}
         + \ a \ \left[{\t {12\over m^2(m+1) }} \alpha_{(1)}^2-
                       {\t {24\over m(m^2-1) }} \alpha_{(2)}\right],&(2.11)\cr
\rho^{(2)}=& a (Da) \left[ D\alpha_{(1)}
           +  {\t {3\over m}}  \alpha_{(1)}^2
           -  {\t {6\over (m-1) }} \alpha_{(2)}\right]\cr
       &+ 2 a^2\left[-{\t {1\over m^2 }} \alpha_{(1)}^3
                    +{\t {3\over m(m-1)}}\alpha_{(1)}\alpha_{(2)} 
                    -{\t {3\over (m-1)(m-2)}}\alpha_{(3)}\right],&(2.12)\cr}$$
while for $m\ge 7$,  
$$\eqalignno{
\rho^{(3)}=& a\ (D^2 a)^2  
        -{\t {60\over m(m+1)(m+3)}}a^2\ D^2a\ D\alpha_{(1)}
        +{\t {1\over 4}} a^{-1} (Da)^4\cr
        &+30 a\ (Da)^2\left[ {\t {(m-1)\over m (m+1)(m+3)}}D\alpha_{(1)}
                              +{\t {1\over m^2 (m+1)}}\alpha_{(1)}^2
                              -{\t {2\over m(m^2-1)}}
        \alpha_{(2)}\right]\cr
        &+{\t {120\over m(m^2-1)(m+3)}} a^2\ Da\Big[ 
            -{\t { (m-1)(m-3)\over m }} \alpha_{(1)} \ D\alpha_{(1)}
            +(m-3) D\alpha_{(2)}\cr
        &  \quad\quad\quad\quad\quad \quad
            -{\t{ (m-1)(2m-3)\over m^2}} \alpha_{(1)}^3
            +{\t {6(m-2)\over m}} \alpha_{(1)}\alpha_{(2)}
            -6 \alpha_{(3)}\Big]\cr
        & +{\t {60\over m(m^2-1)(m+3)}} a^3\Big[ 
             {\t {(m-1)\over m}} (D\alpha_{(1)})^2
            -{\t {4\over m}} D\alpha_{(1)} \ \alpha_{(2)}
            +{\t {(m-1)(2m-3)\over m^3}} \alpha_{(1)}^4 \cr
        &\quad\quad\quad\quad\quad\quad
            -{\t 4 {(2m-3)\over m^2}} \alpha_{(1)}^2\alpha_{(2)}
            +{\t {8\over m}} \alpha_{(1)}\alpha_{(3)}
            +{\t {4\over m}} \alpha_{(2)}^2      
            -{\t {8\over (m-3)}} \alpha_{(4)}    
             \Big].&(2.13)\cr
}$$

\vskip 0.5cm
\vfill
\eject
\noindent
{\bf 3. Top four terms in  the derivatives of differential functions}

In this section we shall obtain 
the expression of $D^k\varphi$ up to top four
highest derivatives. 
Let $\v$ and $u$ be as in Eq.(2.1), and $D^k\v$ be given by (2.3).  
We start by writing  the
expressions of $D^k\v$ for $k\le  6$ up to top four derivatives.
As there is a truncation in the summations, for $k\ge 4$  we should assume that
$n\ge 3 $. 
$$\eqalignno{
D\v=&\sum_{i=0}^n \v_i \ u_{i+1}\ + \ \v_x,&(3.1a)\cr 
D^{2}\v=& \sum_{i=0}^n \v_i\ u_{i+2} +\sum_{i=0}^n D\v_i\ u_{i+1}
+D\v_x, &(3.1b)\cr
D^{3}\v=& \sum_{i=0}^n \v_i\ u_{i+3} +2 \sum_{i=0}^n D\v_i\ u_{i+2}
+\sum_{i=0}^n D^2\v_i \ u_{i+1}+
D^2\v_x,&(3.1c)\cr
 D^{4}\v=& \sum_{i=n-3}^n \v_i\ u_{i+4} +3 \sum_{i=0}^n D\v_i\ u_{i+3}
+3 \sum_{i=0}^n D^2\v_i \ u_{i+2}
+\sum_{i=0}^nD^3\v_i\ u_{i+1}
+D^3\v_x+ O(u_{n}),&(3.1d)\cr
D^{5}\v=& \sum_{i=n-3}^n \v_i\ u_{i+5} +4 \sum_{i=n-2}^n D\v_i\ u_{i+4}
+6 \sum_{i=0}^n D^2\v_i \ u_{i+3}
+4\sum_{i=0}^n D^3\v_i\ u_{i+2}\cr
&+\sum_{i=0}^n D^4\v_i \ u_{i+1}
+D^4\v_x+ O(u_{n+1}),&(3.1e)\cr
 D^{6}\v=& \sum_{i=n-3}^n \v_i\ u_{i+6} 
+5 \sum_{i=n-2}^n D\v_i\ u_{i+5}
+10 \sum_{i=n-1}^n D^2\v_i \ u_{i+4}
+10\sum_{i=0}^n D^3\v_i\ u_{i+3}\cr
&+5\sum_{i=0}^n D^4\v_i \ u_{i+2}
+\sum_{i=0}^n D^5\v_i \ u_{i+1}
+D^5\v_x+ O(u_{n+2}).&(3.1f)\cr
}$$
For  $k\ge 7$, we keep terms of  orders $u_{n+k}$, $u_{n+k-1}$,
$u_{n+k-2}$ and $u_{n+k-3}$ and write the general formula as below.
$$ \eqalignno{
D^{k}\v=& 
\v_n \ u_{n+k}+\v_{n-1}\  u_{n+k-1} +\v_{n-2}\  u_{n+k-2} 
+\v_{n-3} \ u_{n+k-3}+\dots\cr
&+(k-1)\left[ 
D\v_n \ u_{n+k-1}
+ D\v_{n-1}\  u_{n+k-2} 
+D\v_{n-2} \ u_{n+k-3} +\dots\right]\cr
&+{\t {k-1\choose 2}}\left[ 
D^2\v_n \ u_{n+k-2}
+ D^2\v_{n-1}\  u_{n+k-3}  +\dots\right]\cr
&+{\t {k-1\choose 3}}\left[ 
D^3\v_n \  u_{n+k-3} +\dots\right]\cr
&+\dots\quad \dots\cr
&+{\t {k-1\choose 2}}
\left[ 
D^{k-3}\v_{n} \ u_{n+3}+
D^{k-3}\v_{n-1}\  u_{n+2}+
D^{k-3}\v_{n-2}\  u_{n+1}+\dots+
D^{k-3}\v_{1}\  u_{4}+
D^{k-3}\v_{0}\  u_{3}\right]\cr
& +(k-1)
\left[ 
D^{k-2}\v_{n}\  u_{n+2}+
D^{k-2}\v_{n-1}\  u_{n+1}+
D^{k-2}\v_{n-2}\  u_{n  }+\dots+
D^{k-2}\v_{1}\  u_{3}+
D^{k-2}\v_{0}\  u_{2}\right]\cr
& +
\left[ 
D^{k-1}\v_{n}\  u_{n+1}+
D^{k-1}\v_{n-1}\  u_{n}+
D^{k-1}\v_{n-2}\  u_{n-1}+\dots+
D^{k-1}\v_{1}\  u_{2}+
D^{k-1}\v_{0}\  u_{1}\right]\cr
& \quad\quad +D^{k-1}\v_x+ O(u_{n+k-4}). &(3.2)  \cr }$$
Note that, for $k\ge 7$, only the indicated terms of the first four
lines contribute to the top four terms.  In the last three lines, 
there are contributions from  
top three  terms in $D^{k-1}\v_i$, 
top two  terms in $D^{k-2}\v_i$ 
and from the top term in $D^{k-3}\v_i$.  
For $k\ge 7$ these different types of contributions do not mix up and
we can obtain a general formula by inspection, obtaining recursively
the  top first, second and third terms of any derivative.  
Once the general formula is ``guessed'' it can be proved easily by
induction.

\proclaim Proposition 3.1. 
Let $\v$ and $u$ be as in Eq. (2.1) and assume that  $n\ge 3$.  
Then 
$$\eqalignno{
D^{k}\v =& \v_n \ u_{n+k}+P(u_{n+k-1}), \quad k\ge 2,&(3.3a)\cr
D^{k}\v =& \v_n\  u_{n+k} +[\v_{n-1}+ k D\v_n ]
           u_{n+k-1}+P(u_{n+k-2}), \quad k\ge 3,&(3.3b)\cr
D^{k}\v =& \v_n \ u_{n+k} +[\v_{n-1}+ k D\v_n
                 ]u_{n+k-1}\cr
         &+ [\v_{n-2} + k \ D\v_{n-1} +{\t {k \choose 2}} D^2\v_n]u_{n+k-2}
                 +P(u_{n+k-3})\quad k\ge 5,&(3.3c)\cr
D^k\v =& \v_n u_{n+k} + [ \v_{n-1} + k \ D\v_n] u_{n+k-1}\cr
       &+ [\v_{n-2} + k \ D\v_{n-1} +{\t{k \choose 2}} D^2\v_n]
                 u_{n+k-2}\cr
       &+[\v_{n-3}+k\ D\v_{n-2} +{\t{k\choose 2}} D^2\v_{n-1}
                  +{\t{k\choose 3}}
                 D^3\v_n]u_{n+k-3}+P(u_{n+k-4}),\quad k\ge 7&(3.3d)
\cr}$$
and the expressions of $D^k\v$ for $k=1,\dots 6$ are
$$\eqalignno{
D\v  =&\v_n \ u_{n+1}+O(u_n),&(3.4a)\cr
D^{2}\v=& \v_n \  u_{n+2} 
+[\v_{n-1} +2 D\v_n]u_{n+1}
-\v_{nn}\ u_{n+1}^2
+O(u_n),&(3.4b)\cr
D^{3}\v=& \v_n \ u_{n+3} 
+[\v_{n-1} +3 D\v_n]u_{n+2}
+[\v_{n-2} +3 D\v_{n-1} +3D^2\v_{n}]u_{n+1}\cr
&\quad
-3\v_{nn}\ u_{n+1}\  u_{n+2}
-3 D\v_{nn}\  u_{n+1}^2 +\v_{nnn}\ u_{n+1}^3\cr
&\quad -3 \v_{n,n-1}\ u_{n+1}^2+O(u_n),&(3.4c)\cr
D^4\v=& \v_n\  u_{n+4} 
+[\v_{n-1} +4 D\v_n]u_{n+3}
+[\v_{n-2} +4 D\v_{n-1} +6 D^2\v_{n}]u_{n+2}\cr
&\quad +[\v_{n-3}+4 D\v_{n-2} +6 D^2\v_{n-1} +4 D^3\v_n] u_{n+1}\cr
&\quad
-4\v_{nn}\ u_{n+1}\  u_{n+3}
-6 D^2\v_{nn}\  u_{n+1}^2 
-12 D\v_{nn}\ u_{n+1}\ u_{n+2}
-3 \v_{nn}\ u_{n+2}^2\cr
&\quad 
+6 \v_{nnn}\ u_{n+1}^2 \ u_{n+2}
-10 \v_{n,n-1}\  u_{n+1}\ u_{n+2}
-12 D\v_{n,n-1}\  u_{n+1}^2\cr
&\quad
+4 D\v_{nnn}\ u_{n+1}^3
-\v_{nnnn}\ u_{n+1}^4
+6 \v_{nn,n-1}\ u_{n+1}^3\cr
&\quad
-4 \v_{n,n-2}\ u_{n+1}^2
-3\v_{n-1,n-1}\ u_{n+1}^2
+O(u_{n}),&(3.4d)\cr
D^{5}\v
=& \v_n\  u_{n+5} 
+[\v_{n-1} +5 D\v_n]u_{n+4}
+[\v_{n-2} +5 D\v_{n-1} +10 D^2\v_{n}]u_{n+3}\cr
&\quad +[\v_{n-3}+5 D\v_{n-2} +10 D^2\v_{n-1} +10 D^3\v_n] u_{n+2}\cr
&\quad
-10 \v_{nn}\ u_{n+2}\ u_{n+3}
-15 D\v_{nn}\  u_{n+2}^2
-10 \v_{n,n-1}\  u_{n+2}^2
+P(u_{n+1}),&(3.4e)\cr
D^{6}\v
=& \v_n\  u_{n+6} 
+[\v_{n-1} +6 D\v_n]u_{n+5}
+[\v_{n-2} +6 D\v_{n-1} +15 D^2\v_{n}]u_{n+4}\cr
&\quad +[\v_{n-3}+6 D\v_{n-2} +15 D^2\v_{n-1} +20 D^3\v_n] u_{n+3}\cr
&\quad
-10 \v_{nn}\ u_{n+3}^2
+P(u_{n+2}).&(3.4f)
\cr}$$

\vskip 0.3cm
\noindent
{\bf Proof.}  
The expressions (3.4a-f) are obtained iteratively 
by  expanding the derivatives in (3.1a-f) and keeping the
relevant terms, which is a  tedious but straightforward computation.

Similarly, to  prove (3.3d) by induction, one has  to check that it holds for
$k=7$, by replacing $k=7$ in (3.2). Then assuming (3.3d) holds for
$k\ge 7$, 
we compute 
$$\eqalignno{
D^{k+1}\v =& \v_n \ u_{n+k+1} + D\v_n\  u_{n+k}\cr
           & +[ \v_{n-1} + k \ D\v_n] u_{n+k}
             +[ D\v_{n-1} + k \ D^2\v_n] u_{n+k-1}\cr
       &+ [\v_{n-2} + k \ D\v_{n-1} +{\t{k \choose 2}} D^2\v_n]
                 u_{n+k-1}\cr
      &+ [D\v_{n-2} + k \ D^2\v_{n-1} +{\t{k \choose 2}} D^3\v_n]
                 u_{n+k-2}\cr
       &+[\v_{n-3}+k\ D\v_{n-2} +{\t{k\choose 2}} D^2\v_{n-1}
                  +{\t{k\choose 3}}
                 D^3\v_n]u_{n+k-2}+P(u_{n+k-3}),&(3.5a)\cr
      =& \v_n \ u_{n+k} +
             [ \v_{n-1} + (k+1) \ D\v_n]  \ u_{n+k}\cr
           &+ \left[\v_{n-2} + (k+1) \ D\v_{n-1} 
            +\left[ {\t{k \choose 2}}+k\right] D^2\v_n\right]
                 u_{n+k-1}\cr
       &+\left[\v_{n-3}+(k+1)k\ D\v_{n-2} +
                   \left[  {\t{k\choose 2}}+k\right] D^2\v_{n-1}
                  +\left[{\t{k\choose 3}}+{\t{k\choose 2}}\right]
                 D^3\v_n
\right] u_{n+k-2}\cr
      &\quad\quad +P(u_{n+k-3}).&(3.5b)
\cr}$$
Using the relations 
$${k \choose 2}+k={k+1\choose 2},\quad\quad 
 {k\choose 3}+{k\choose 2}={k+1\choose 3}$$\
 in (3.5b), we
obtain Eq.(3.3d).  The validity of the remaining expressions in
(3.3a-d) for the indicated values of $k$  can
be checked from Eqs.(3.4a-f).
\hfill $\bullet$

\vskip 0.5cm
\noindent
{\bf 4. General Results on Classification}
\vskip .2cm

In this section we shall obtain certain general results on
classification. 
We start by stating the following well known result which is proved
easily from the conserved density condition.

\proclaim Proposition 4.1.  Let $F=F(x,t,u,\dots,u_{2k})$.  If
$\rho=\rho(x,t,u,\dots,u_n)$ is a conserved density of the evolution
equation $u_t=F$, then $\rho_{nn}=0$, for $n\ge 2$ and $k\ge 1$.

\noindent
{\bf Proof.} Let $n=2k+l$. Then $D_t\rho$ is the sum of terms
$\rho_{2k+l-j}D^{2k+l-j}F$. Integrating this  $k-j$ times we obtain 
$D^{k-j}\rho_{2k+l-j}D^{k+l}F$ which is quadratic in
$u_{3k+l-j}$. Thus the contribution to the highest order nonlinearity comes
from the first term corresponding to $l=0$ only. It follows that
$\rho_{nn}=0$.
\hfill $\bullet$  

\vskip 0.3cm
For $m$ odd,  top $2$ terms in the expansion of the derivatives
contribute to the highest order nonlinearity and we have a nontrivial result.

\proclaim Proposition 4.2. Let $\rho$ be a conserved density of order 
$n\ge m$ for
an evolution of order $m=2k+1$. Then for $k\ge 2$,
$${\t m\over 2} \ F_m\ D\rho_{nn}-(n-{\t 1\over 2}m)\ DF_m\ \rho_{nn}=
  \rho_{nn} F_{m-1}\quad \quad m\ge 7,\quad n\ge m\eqno(4.1)$$

\noindent
{\bf Proof.} From the proof of Proposition 5.1 
it will be seen that only top terms in $D_t\rho$  will contribute to
the highest order nonlinearity. Let $n=2k+l+1$.  Then the
highest nonlinearity will be of order  $3k+l+1$ and 
$$\eqalignno{D_t \rho\cong &\rho_nD^{2k+l+1}F+\rho_{n-1}D^{2k+l}F+
O(u_{3k+l}^2)\cr
\cong&(-1)^{k+1}\left[ D^{k+1}\rho_n -D^{k}\rho_{n-1} \right]
D^{k+l}F +O(u_{3k+l}^2).\cr
}$$
Furthermore, only top two derivatives of each term will contribute 
to the highest order nonlinearity.  Hence we can use the formula
(3.3b) which is valid beyond the third derivative.  We note that (3.3b)
can be used for $k=2$ also, because only the top derivative in $D^k\rho_{n-1}$ 
is needed. Thus 
$$\eqalignno{(-1)^{k+1}\ D_t \rho\cong &
\Big[ \rho_{nn}u_{3k+l+2}+(k+1) D\rho_{nn} u_{3k+l+1}\Big]\cr
&\quad\quad \times \Big[ F_mu_{3k+l+1} +[F_{m-1}+(k+l)DF_m]u_{3k+l}\Big]
\ u_{3k+l+1}^2
+O(u_{3k+l}^2)\cr
\cong&\Big[-{\t {1\over 2} } D(\rho_{nn}F_m)
-\rho_{nn}[F_{m-1}+(k+l)DF_m]+(k+1)F_mD\rho_{nn}\Big]
\ u_{3k+l+1}^2
+O(u_{3k+l}^2)\cr
\cong& \left[(k+{\t{1\over 2}}) D\rho_{nn}F_m -(k+l+{\t{1\over
2}})\rho_{nn}DF_m-\rho_nnF_{m-1}\right]
\ u_{3k+l+1}^2
+O(u_{3k+l}^2).\cr
}$$
The coefficient of $\ u_{3k+l+1}^2$ gives (4.1) and the proposition is
proved. \hfill$\bullet$.
\vskip 0.3cm

\noindent
The implications of (4.1) are listed below. 
Conserved densities of order $n>m$ are necessarily quadratic in
the highest derivative and there is essentially only one conserved
density at each order.  
The corollary can
be proved easily by replacing the indicated values for $n$ in (4.1).

\proclaim Corollary 4.3.  Let $\rho$ be a conserved density of order $n\ge
m$ for an evolution equation of order $m=2k+1\ge 5$.  Then
\item{i.} For $n=m$, 
$$ 
F_m\ \rho_{mmm}=F_{mm}\ \rho_{mm}.\eqno(4.2a)$$
\item{ii.} If $\rho$ and $\eta$ are two conserved densities of order
$n$ with $\rho_{nn}\ne 0$, $\eta_{nn}\ne 0$, then 
$${D\rho_{nn}\over \rho_{nn}}= {D\eta_{nn}\over \eta_{nn}}\eqno(4.2b)$$  
\item{iii.} For $n>m$, 
 $$ \rho_{nnn}=0.\eqno(4.2c)$$

Applying (4.2a) to $\rho^{(-1)}$, we will show that
there are three classes for $F$, the first one being the quasilinear equations.
  For $ m=3$, both  the essentially nonlinear classes  are
well known candidates for integrable equations.

\proclaim Proposition 4.4.  Let $\rho=F_m^{-1/m}$ be a conserved
 density of the evolution equation $u_t=F$.   Then $F$ belongs to one
 of the classes below.
$$\eqalignno{
&i.   \quad \rho_m=0:\quad\quad F=Au_m+B,\quad\quad A_m=B_m=0, &(4.3a)\cr
&ii.  \quad \rho_{mm}=0:\quad F=(Au_m+B)^{1-m}+C,\quad\quad 
                         A_m=B_m=C_m=0,&(4.3b)\cr
&iii. \quad \rho_{mm}\ne 0: \quad 
  F_m= \left[ c^{(1)} F^2 +c^{(2)}F +c^{(3)}\right]^{m/(m-1)},\quad
c^{(1)}_m=c^{(2)}_m=c^{(3)}_m=0.
&(4.3c)
}
$$

\vskip 0.3cm
\noindent
{\bf Proof.} First note that $\rho_m=0$ 
corresponds to $F_{mm}=0$, in which case the conserved density
condition (4.2a) is identically satisfied and gives the class (4.3a). 
Then we compute 
$$\rho_{mm}={\t{1\over m}}({\t {1\over m}}+1) 
F_m^{-{1\over m}-2}F_{mm}^2
-{\t {1\over m}} F_m^{-{1\over m}-1} F_{mmm}\eqno(4.4)$$
and $\rho_{mm}=0$ gives
$$ ({\t{1\over m}}+1) F_{mm}^2+F_m F_{mmm}=0\eqno(4.5)$$
which can be integrated to give  the class of equations (4.3b).
For $\rho_{mm}\ne 0$, Eq.(4.2a) gives
$${F_{mm}\over F_m}={\rho_{mmm}\over \rho_{mm}},\eqno(4.6)$$
which implies that $F_m$ is proportional to $\rho_{mm}$ and using
Eq.(4.4) we obtain a third order ordinary differential equation.
Integrating this twice we obtain 
$$\rho_m=\mu F+\nu,\quad \mu_m=\nu_m=0.\eqno(4.7)$$
Substituting $\rho=F_m^{-1/m}$ in (4.7), integrating once more and
renaming integration constants  we have Eq.(4.3c). We also note that
the class (4.3c) with $c^{(1)}=0$ corresponds to the class (4.3b).
\hfill $\bullet$

\vskip 0.3cm
\noindent
The equation (4.3c) can be integrated for $m=3$ but it has  no
antiderivative for $m\ge 5$.  Thus for $m=3$ we have the following
well known result.

\proclaim Corollary 4.5 (Third order equations).
  Let  $u_t=F(x,t,u,\dots,u_3)$.  If $\rho=F_3^{-1/3} $ is a conserved
 density  then $F$ belongs to one
 of the classes below.
$$\eqalignno{
&\quad F=Au_3+B,\quad\quad A_3=B_3=0, &(4.8a)\cr
&\quad F=(Au_3+B)^{-2}+C,\quad\quad A_3=B_3=C_3=0,&(4.8b)\cr
&\quad F=
{2A u_3+B\over
(A u_3^2 +B u_3 +C)^{1/2}}   
+E,\quad A_3=B_3=C_3=E_3=0.&(4.8c)\cr
}
$$

\vskip 0.3cm
We now arrive at the discussion of equations of order $m\ge 5$. 
Recall that the expression of $\rho^{(1)} $ given by Eq.(2.11)
which is valid for $m\ge 5$ is of the form
(4.2c) with 
$$\rho_{nn}=2 a^{-1} a_m^2,\eqno(4.9)$$
where $n=m+1$.  From (4.1) we obtain 
$${a_{mm}\over a_m}-{\t m+3\over 2}{a_m\over a}=0\eqno(4.10)$$
which gives 
$$a=(Au_m+B)^{-{2\over m+1}}\eqno(4.11)$$
where $A$ and $B$ are independent of $u_m$.  It follows that 
$$F={1\over A} {1+m\over 1-m} (A u_m+B)^{(1-m)/(1+m)}+C,\quad A_m=B_m=C_m=0.
\eqno(4.12)$$
It can be checked that this corresponds to 
$$c_1={\t \left({1-m\over 1+m}\right)^2}A^2,\quad c_2=-2c_1C,\quad
c_3=c_1C^2\eqno(4.13)$$
in (4.3c) and it is inconsistent with (4.3b).
As we shall see that for $m\ge 7$ only quasilinear equations are
allowed, we state the Proposition below for $m=5$.

\proclaim Proposition 4.6 (Fifth order equations).  
Let $\rho=\rho^{(1)} $ given by Eq.(2.11)
be a conserved density of the evolution equation
$u_t=F(x,t,u,\dots,u_5)$. 
Then integrable equations consist of two classes
$$\eqalignno{
&\quad F=Au_5+B,\quad\quad A_5=B_5=0, &(4.14a)\cr
&\quad 
F=-{3\over 2A} (A u_1+B)^{-2/3}+C,\quad
A_5=B_5=C_5=0,\quad 
A_4=C_4=0.
&(4.14b)\cr
}
$$

\vskip 0.3cm
\noindent
{\bf Proof.}
Substituting  (4.12) in the conserved density condition  and
taking logarithmic derivatives we obtain
$$F_{m-1}/F_m=m DA/A.\eqno(4.15)$$
Taking the derivative of $F$ with respect to $u_{m-1}$ and
substituting, we see that the expression above should be linear in
$u_m$.  The second derivative with respect  to $u_m$ gives
$C_{m-1}=0$, while the coefficient of the linear term leads to
$A_{m-1}=0$. Substituting $m=5$,  we have (4.14b).
\hfill$\bullet$

\vskip 0.3cm
\noindent
{\bf Remark 4.7.} Let $\rho$ be a conserved density of order $6$ for
an evolution equation of order $5$. In the  expression $D_t\rho$,
the coefficients of 
$u_8^2u_6$, $u_7^3u_6$ and  $u_7^2u_6^3$
depend only on $\rho_{66}$. These expressions give   respectively
$$\eqalignno{
&  7   F_{55}\ \rho_{66}
  -5  \rho_{665}=0,\cr
&  -42 F_{555}\ \rho_{66}
  -  2 F_{55 }\ \rho_{665}
  +25 F_{5 }\ \rho_{6655}=0,\cr
&  -28 F_{5555}\ \rho_{66}
   +24 F_{555 }\ \rho_{665}
   -23 F_{55} \ \rho_{6655}
  +15 F_{5 }\ \rho_{66555}=0.\cr
}$$
After eliminations we obtain
$$\eqalignno{
   F_{55}\ \rho_{66}  =&{\t {5\over 7}} F_5\ \rho_{665}=0,&(4.16a)\cr
   F_{55}\ \rho_{665} =&{\t {1\over 2}} F_5\ \rho_{6655}=0,&(4.16b)\cr
   F_{55}\ \rho_{6655}=&{\t {5\over13}} F_5\ \rho_{66555}=0.&(4.16c)\cr
}$$
It can be checked that the equations above are consistent with 
$$F_5=a^5,\quad \rho_{66}=a^{-1} \ a_5^2, \quad a_{55}=4 \ a^{-1} \
a_5^2,$$
where the last equation corresponds to Eq.(4.10) with $m=5$. Thus  we
cannot exclude the possibility of the 
existence of essentially nonlinear equations at the $5$th order

\vskip 0.5cm
\noindent
{\bf 5. Equations of order greater than seven.}
\vskip 0.3cm

We consider now evolution equations of  order $m\ge 7$ admitting a
conserved density $\rho$ of order $n=m+1$, i.e.,
$$m=2k+1,\quad n=m+1=2k+2.$$
From Corollary 4.3, we know that $\rho$ is quadratic in $u_n$, i.e,
$$\rho=\alpha u_n^2+\beta u_n+\gamma, \quad\quad
\alpha_n=\beta_n=\gamma_n=0.
\eqno(5.1)$$
For convenience, in the proof of Lemma  5.3 and afterwards,
 we shall use the notation 
$$
\rho'=\rho_{m+1},\quad 
\rho''=\rho_{m+1,m+1}, \quad 
\rho'_j=\rho_{m+1,j},\quad 
\rho''_j=\rho_{m+1,m+1,j},\quad  
j=0,\dots  m.$$
The following relation can be easily deduced from Eq.(5.1).

\proclaim Lemma 5.1. 
Let $\rho=\rho(x,t,u,\dots,u_{m+1}) $ 
be a quadratic function of $u_{m+1}$.  Then
$$\eqalignno{
D\rho'=&\rho''_m\  u_{m+1} + O(u_m),\quad
D^2\rho''=\rho''_m\ u_{m+2} + P(u_{m+1}),\quad
D^3\rho''=\rho''_m\ u_{m+3} + P(u_{m+2}),\cr
D\rho'_j  =&\rho''_j\ u_{m+2} + P(u_{m+1}),\quad
D^2\rho'_j=\rho''_j\ u_{m+3} + P(u_{m+2}).&(5.2)}$$

We shall now show that, in the integral of $D_t\rho$, 
the contribution to top two nonlinear terms
$u_{3k+2}^2$ and $u_{3k+1}^2$ will come from top four derivatives.


\proclaim Proposition 5.2.
Let $\rho=\rho(x,t,u,\dots u_n)$ and $u_t=F(x,t,u,\dots,u_m)$, where
$n=m+1$ and $m=2k+1$.  Then
$$(-1)^{k+1}D_t\rho\cong 
   \Big[D^{k+1}\rho_n  - D^k\rho_{n-1}\Big]\ D^{k+1}F 
- \Big[D^k\rho_{n-2}\ D^kF- D^{k-1}\rho_{n-3}\Big]\  D^k F
+O(u_{3k}).\eqno(5.3)$$

\noindent
{\bf Proof.} We write
$$\eqalignno{
D_t\rho=& 
\rho_n \ D^n F 
+ \rho_{n-1}\  D^{n-1}F 
+\rho_{n-2}\ D^{n-2}F
+\rho_{n-3}\ D^{n-3}F
+\rho_{n-4}\ D^{n-4}F
+\dots 
+\rho_0\ F
+\rho_t,\cr
=&
\rho_{2k+2}\  D^{2k+2} F 
+ \rho_{2k+1}\  D^{2k+1}F 
+\rho_{2k}\ D^{2k}F
+\rho_{2k-1}\ D^{2k-1}F
+\rho_{2k-2}\ D^{2k-2}F
+\dots\cr 
&+\rho_0\ F
+\rho_t,\cr
}$$
We integrate the first term $k+1$ times, the second and the third $k$
times etc., so that each term is a product of terms of orders equal or
differing at most by one. Hence we have 
$$\eqalignno{
(-1)^{k+1} D_t\rho\cong & 
 \Big[ D^{k+1}\rho_{2k+2}- D^k    \rho_{2k+1}\Big] \ D^{k+1}F
-\Big[ D^k    \rho_{2k}  - D^{k-1}\rho_{2k-1}\Big]\  D^k F\cr
& 
- \Big[ D^{k-1}\rho_{2k-2}- D^{k-2}\rho_{2k-3}\Big] \ D^{k-1}F
+ \dots  +\rho_0\ F+\rho_t.\cr
}$$
It will be sufficient to show that third term 
is of order $u_{3k}$, because by a similar reasoning it can be seen
that the remaining terms are of lower orders. 
As $\rho_{2k-2}$ is of polynomial in $u_{2k+2}$ the term in the
brackets is linear in $u_{3k+1}$  and polynomial in other derivatives
of order larger than $m$.  Similarly  $D^{k-1}F$ is polynomial in
$u_{3k}$, hence after integration by parts, their product is of order
$u_{3k}$.
It follows that 
the contribution to top two nonlinear terms comes from top four terms 
in the expansion of $D_t \rho$.\hfill$\bullet$
\vskip 0.3cm

The contribution to the coefficient of  $u_{3k+2}^2$
comes from the top two derivatives of the first term of (5.3) and it 
has already been given in (4.1).  We shall  need
now the coefficients of 
$$u_{3k+2}^2\ u_{2k+2}\quad \quad {\rm and} \quad \quad  
  u_{3k+1}^2\  u_{2k+4}.$$
We shall now show that at most top four terms of each derivative
contribute to the coefficient of $u_{3k+1}^2$ and actually the
contribution to the  coefficient of $u_{3k+1}^2 u_{2k+4}$ comes only
from the top two terms in each product in 
$$(D^{k+1}\rho_n-D^k\rho_{n-1})D^{k+1}F.$$
For this we shall compute the relevant terms in each factor of
 $D_t\rho$ separetely.

\proclaim Lemma 5.3.
Let $\rho0\rho(x,t,u,\dots,u_n)$ be  quadratic in  $u_n=u_{m+1 }$, 
$F=F(x,t,u,\dots,u_m)$ and   let
$k\ge 6$. 
Then
$$\eqalignno{
D^{k+1}\rho'-D^k\rho_m=&
\rho'' \ u_{m+k+2} 
+\left[ r_1\ \rho''_m \ u_{m+1} + O(u_m)\right]\ u_{m+k+1} \cr
&\quad +\left[ r_2\ \rho''_m \ u_{m+2} + P(u_{m+1})\right]\ u_{m+k}\cr
&\quad
+\left[ r_3\ \rho''_m \ u_{m+3} + P(u_{m+2})\right]\ u_{m+k-1} \cr
&\quad + P(u_{m+k-2})&(5.4a)\cr
D^{k}\rho_{m-1}-D^{k-1}\rho_{m-2}=&
P(u_{m+1})\ u_{m+k+1}+P(u_{m+2})\ u_{m+k} +P(u_{m+k-1})&(5.4b)\cr
D^{k+1}F=&
F_{m}  \ u_{m+k+1} 
+\left[ f_1\ F_{mm}  \ u_{m+1} + O(u_m)\right]\ u_{m+k}\cr
&\quad +\left[ f_2\ F_{mm} \ u_{m+2} + P(u_{m+1})\right]\ u_{m+k-1}\cr
&\quad +\left[ f_3\ F_{mm} \ u_{m+3} + P(u_{m+2})\right]\ u_{m+k-2}\cr
&\quad + P(u_{m+k-3})&(5.4c)\cr
D^kF=&O(u_m)\ u_{m+k}+ P(u_{m+1}) \ u_{m+k-1} + P(u_{m+k-2})&(5.4d)\cr
}$$
where 
$$\eqalignno{
r_1=&k+1,\quad   
r_2=1+{\t{k+1\choose 2} },\quad 
r_3={\t{k+1\choose 2   }}+{\t{k+1\choose 3   }}-{\t{k\choose 2
}}&(5.5a)\cr
f_1=&k+1,\quad   
f_2={\t{k+1\choose 2} },\quad 
f_3={\t{k+1\choose 3}}&(5.5b)\cr
}
$$

\noindent
{\bf Proof.} 
We substitute $\v= \rho'$ in (3.3d) which is valid for $k+1\ge 7$ and get
$$\eqalignno{
D^{k+1}\rho'=&\rho'' \ u_{m+k+2} +[\rho'_m+(k+1)D\rho'']u_{m+k+1}\cr
&\quad +[\rho'_{m-1} +(k+1) D\rho'_m+{\t {k+1\choose 2}}D^2\rho'']\ u_{m+k}\cr
&\quad +[\rho'_{m-2} +(k+1) D\rho'_{m-1}+{\t {k+1\choose 2}}D^2\rho'_m
   +{\t{k+1 \choose 3 }}D^3\rho'' ] \ u_{m+k-1}\cr
&\quad+ P(u_{m+k-2})\cr
=&\rho'' \ u_{m+k+2} 
    +[\rho'_m+(k+1)\rho''_m  u_{m+1} +O(u_m) ]u_{m+k+1}\cr
&\quad +\Big[[(k+1)+{\t {k+1\choose 2 }}]\rho''_m u_{m+2} 
+P(u_{m+1})\Big]\ u_{m+k}\cr
&\quad +\Big[[ {\t{k+1\choose 2} }+{\t {k+1\choose 3 }}]\rho''_m u_{m+3} 
+P(u_{m+2})\Big]\ u_{m+k-1}\cr
&\quad+ P(u_{m+k-2}).&(5.6)\cr
}$$
Then we repeat the same computations for $D^k\rho_m$ up to top three
terms using (3.3c) which is valid for $k\ge 5$. 
$$\eqalignno{
D^{k}\rho_m=&\rho'_m \ u_{m+k+1} +[\rho_{mm}+kD\rho'_m]u_{m+k}\cr
&\quad +[\rho_{m,m-1} +k D\rho_{mm}+{\t {k\choose 2}}D^2\rho'_m]\ u_{m+k-1}\cr
&\quad+ P(u_{m+k-2})\cr
=&\rho'_m \ u_{m+k+1} 
    +[k\rho''_mu_{m+2} +P(u_{m+1}) ]u_{m+k}\cr
&\quad +\Big[ {\t {k\choose 2 }}\rho''_m u_{m+3} +P(u_{m+2})\Big]
\ u_{m+k-1}\cr
&\quad+ P(u_{m+k-2}).&(5.7)\cr
}$$
Subtructing (5.7) from (5.6) we obtain (5.4a).  The derivation of
Eqs.(5.4b-d) is straightforward. \hfill $\bullet$

\vskip 0.3 cm

In the products of (5.4a) with (5.4c) and (5.4b) with (5.4d) most of
the terms will not contribute to the top two nonlinear terms, as
indicated below.

\proclaim Lemma 5.4. For $k\ge 6$, the following relations hold.
$$\eqalignno{
&u_{m+k+2}\ P(u_{m+k-3})\cong
u_{m+k+1}\ P(u_{m+k-2})\cong 
u_{m+k}  \ P(u_{m+k-1})\cong 
P(u_{m+k-1}) &(5.8a)\cr
&u_{m+k+2}\ u_{m+k-1} \  P(u_{m+1})\cong 
u_{m+k+2}\ u_{m+k-2} \  P(u_{m+2})\cr
& \quad\quad \cong 
u_{m+k+1}\ u_{m+k  } \  P(u_{m+1})\cong 
u_{m+k+1}\ u_{m+k-1} \  P(u_{m+2})\cong 
u_{m+k}^2\   P(u_{m+2})&(5.8b)\cr
&u_{m+k+2}\  u_{m+k} \O(u_m)\cong 
u_{m+k+1}^2\ O(u_m)&(5.8c)\cr
}$$

\vskip .3cm
\noindent
It follows that the contribution to the top two nonlinear terms of
$D_t\rho$ comes from 
$$\eqalignno{
\Big[
\rho'' \ u_{m+k+2}& 
+\rho''_m \left[ 
   r_1\  u_{m+1} \ u_{m+k+1}
+  r_2\  u_{m+2} \ u_{m+k}
+  r_3\  u_{m+3} \ u_{m+k-1}\right]\Big]\cr
&\times \Big[
F_{m}  \ u_{m+k+1} 
+F_{mm} \left[ 
  f_1\ u_{m+1} \  u_{m+k}
+ f_2\ u_{m+2} \ u_{m+k-1}
+ f_3\ u_{m+3} \ u_{m+k-2}\right]\Big],&(5.9) \cr
}$$
where $r_i$'s and $f_i$'s are given by Eqs.(5.5a-b).
Substituting these values and integrating by parts we obtain the
   following result.

\proclaim Proposition 5.5.  Let $u_t=F(x,t,u,\dots,u_m)$ 
be an evolution equation of order
$m$ and  $\rho=\rho(x,t,u,\dots,u_n)$ 
be a conserved density of order $n=m+1$ satisfying $\rho_{nnn}=0$.  
The coefficients of 
$u_{3k+2}^2u_{2k+2}$ and $u_{3k+1}^2u_{2k+4}$
give respectively
$$\eqalignno{ &
 (2k+1)\rho''_m F_m=(2k+3)\rho'' F_{mm},&(5.10a)\cr
&(2k+1)(k^2+k+6)\rho''_m F_m= (2k+3)(k+1)(k+2)\rho'' F_{mm}&(5.10b)\cr
}$$
where $\rho''=\rho_{nn}$

\noindent
{\bf Proof.} The general formulas can be used for $k\ge 6$, i.e.,
$m\ge 13$ and  Eqs.(5.10a-b) can be obtained from Eq.(5.9).  For
$m=5$, straightforward computation of the conserved density condition
is possible with REDUCE and the results have already been given by 
Eqs.(4.161-c).
For $m=7$ and $m=$ it is necessary to use a combination of 
symbolic and analytic computations.
For $m=11$, there is little  discrepency from the general formulas 
and the required coefficient scan be obtained easily with analytical.
 \hfill$\bullet$

\vskip 0.3cm
It can be easily checked that the equations (5.10a) and (5.10b) are 
inconsistent
except for $k=2$.
 For $k=2$, (5.10b) is the coefficient of $u_7^2u_8$,
and leads to Eq.(4.16b) after integration. Thus for $k\ge 3$
 Thus
$$\alpha_m\ F_m= \alpha\ F_{mm}=0.$$
It follows that for $\alpha \ne 0$, $F_{mm}=0$, and we have the
following corollary.

\proclaim Corollary  5.6.  Let $u_t=F(x,t,u,\dots,u_m)$ 
be an evolution equation of order
$m\ge 7$ and  $\rho=\rho(x,t,u,\dots,u_n)$
be a conserved density of order $n=m+1$, with $\rho_{nn}\ne 0$. Then
$F_{mm}=0$.

Note that in order to prove the quasilinearity result, the explicit 
form of $\rho^{(1)}$ is not needed. We have only used here the fact
that it is indeed quadratic in $u_{m+1}$.

\vskip 1cm

{\bf References.}
\baselineskip=12pt
\vskip .1cm
\vskip .1cm

\item{[1]}
J.A. Sanders and J.P. Wang, 
`` On the integrability of homogeneous scalar evolution eqautions'',
{\it Journal of Differential Equations}, vol. 147,(2), pp.410-434,
(1998).
\vskip .1cm

\item{[2]}
J.A. Sanders and J.P. Wang, 
`` On the integrability of non-polynomial scalar evolution eqautions'',
{\it Journal of Differential Equations}, vol. 166,(1), pp.132-150,
(2000).
\vskip .1cm

\item{[3]} A.V. Mikhalov, A.B. Shabat and V.V Sokolov. 
``The symmetry approach
to the classification
of integrable equations" in `{\it What is Integrability?} edited by
V.E. Zakharov (Springer-Verlag, Berlin 1991).
\vskip .1cm

\item{[4]}
R.H. Heredero, V.V. Sokolov and S.I. Svinolupov,
``Classification of 3rd order integrable evolution equations'',
{\it Physica D}, vol.87 (1-4), pp.32-36, (1995).
\vskip .1cm

\item{[5]} P.J. Olver, Lie {\it Application of Lie
Groups to Differential Equations} (Springer-Verlag, Berlin 1993)
\vskip .1cm

\item{[6]} 
A.H. Bilge,
``A REDUCE program for the integration of differential polynomials'', 
{\it Computer Physics Communications}, vol. 71, p.263, (1992).
\vskip 0.1cm


%
%
%
%





\vfill
\eject

\end